# Multibeam Free Space Optics Receiver Enabled by a Programmable Photonic Mesh

Maziyar Milanizadeh[1], SeyedMohammad SeyedinNavadeh[1], Francesco Zanetto[1], Vittorio Grimaldi[1], Christian De Vita[1], Charalambos Klitis[2], Marc Sorel[2], Giorgio Ferrari[1], David A.B. Miller[3], Andrea Melloni[1], Francesco Morichetti[1]

[1] Department of Electronics, Information and Bioengineering, University Politecnico di Milano, Piazza Leonardo da Vinci, 32, 20133, Milano, Italy
[2] School of Engineering University of Glasgow, Glasgow, G12 8QQ, U.K
[3] Ginzton Laboratory, Stanford University, Spilker Building, Stanford, CA 94305, USA

email: francesco.motichetti@polimi.it

*Abstract*— Free-space optics (FSO) is an attractive technology to meet the ever-growing demand for wireless bandwidth in next generation networks. To increase the spectral efficiency of FSO links, transmission over spatial division multiplexing (SDM) can be exploited, where orthogonal light beams have to be shaped according to suitable amplitude, phase, and polarization profiles. In this work, we show that a programmable photonic circuits, consisting of a silicon photonic mesh of tunable Mach-Zehnder Interferometers (MZIs) can be used as an adaptive multibeam receiver for a FSO communication link. The circuit can self-configure to simultaneously receive and separate, with negligible mutual crosstalk, signals carried by orthogonal FSO beams sharing the same wavelength and polarization. This feature is demonstrated on signal pairs either arriving at the receiver from orthogonal directions (direction-diversity) or being shaped according to different orthogonal spatial modes (mode-diversity), even in the presence of some mixing during propagation. The performance of programmable mesh as an adaptive multibeam receiver is assessed by means of data channel transmission at 10 Gbit/s a wavelength of 1550 nm, but the optical bandwidth of the receiver (> 40 nm) allows its use at much higher data rates as well as in wavelength-division multiplexing (WDM) – SDM communication links.

## I. Introduction

The concepts of space diversity and space multiplexing are well established in communications systems and are widely employed in microwave wireless systems to implement high capacity multiple input - multiple output (MIMO) links. In the optical domain, space division multiplexing (SDM) has been known for several decades [1], but only recently it has started to be seriously considered as a strategy to face the capacity crunch of optical fibers [2]. Fiber optic SDM systems exploit multicore or few-mode fibers to increase the spectral efficiency (in terms of bit/Hz/s) of the transmitted signal, but the price to be paid at the receiver is the need for coherent detection assisted by electronic digital signal processing (DSP). Such a DSP should run at the bit rate to recover the signal integrity by undoing the mode mixing occurred during fiber propagation. To reduce the power consumption and the speed limitation of the DSP, several solutions have been proposed to perform all-optical demultiplexing and unmixing of the optical modes at the receiver [3], [4], [5], [6], [7]

The same evolution that fiber optic communications had experienced towards SDM systems is now happening in free space optics (FSO). FSO communications is attracting renewed and ever-increasing interest because it is a potential solution to meet the growing demand for wireless bandwidth and the low latency requirements of Internet of Things (IoT) technologies in next generation networks [8]. As in the case of fiber optic communications, SDM in FSO requires the use of orthogonal sets of beams (or modes), and several pioneering demonstrations have been achieved by using orbital angular momentum (OAM) modes [9] [10], Bessel beams [11], and Laguerre-Gauss modes [12]. To generate a number of orthogonal beam configurations, and to demultiplex them at the receiver, the light beams have to be shaped according to suitable amplitude, phase, and polarization profiles [13]. Traditionally, these operations are performed by using bulk optics, such as classical lenses and diffractive elements. To have more flexibility and reconfigurability in beam manipulation, spatial light modulators (SLMs) are also available [14]; however, they have some limitations, such as a relatively low speed (a few hundred Hz), the possibility of modifying only the phase of light (only a few examples with amplitude control [15], with efficiency penalties) and the need for computationally heavy calibration techniques.

A powerful alternative technology for the manipulation of the FSO beams is offered by programmable photonic integrated circuits (PICs). These circuits are general-purpose meshes of tunable integrated interferometers, arranged either in feed-forward or recursive topologies, which can implement arbitrary linear transformations [16]. Because of their remarkable flexibility, they have been already used in many different applications, including reconfigurable filters [17], unmixing of guided modes [4], vector-matrix multiplication and computing [18], quantum information processing [19] and neural networks [20], [21]. Recently, we employed a silicon photonic mesh of Mach-Zehnder interferometers (MZIs) in a FSO system to control the complex field radiated and captured by an array of optical antennas, demonstrating several functionalities like the generation of perfectly shaped beams with nonperfect optical antennas, imaging through a diffusive medium, and the identification of an unknown obstacles. Significantly, the MZI mesh can self-configure and self-control through simple automated control strategies without the need for global multivariable optimization techniques.

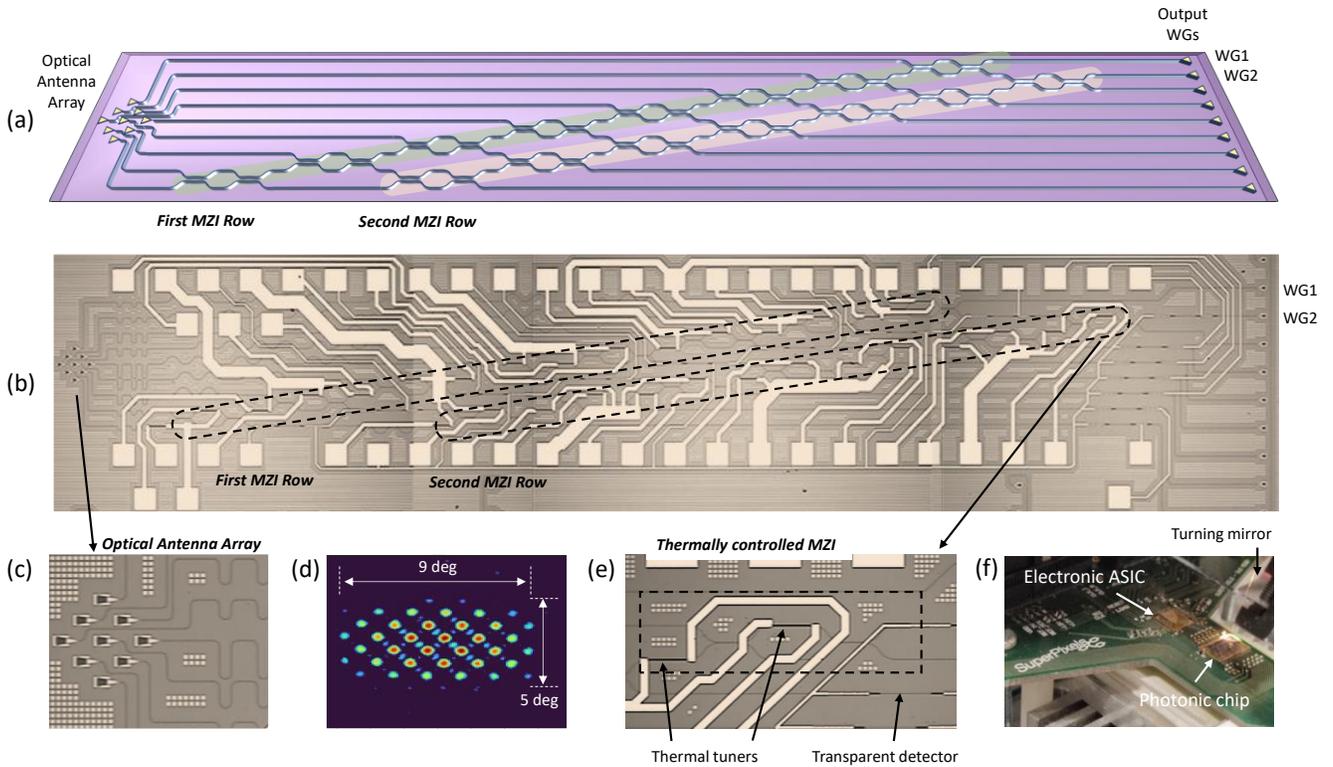

Figure 1: Multi-beam FSO Receiver. (a) Schematic of a 9×2 diagonal mesh comprising two rows of tunable MZIs and implementing a two-beam FSO receiver. The 2D optical antenna array is used to couple free space beams into the silicon waveguides, while the output ports WG1 and WG2 are used to couple the light out to a pair of optical fibers. (b) Microscope picture of the fabricated silicon chip. (c) Detail of the 2D optical antenna array made of 3×3 grating couplers in a square configuration. (d) Measured far-field pattern radiated by the 2D optical antenna array when all the grating couplers are excited with same amplitude and phase. Multiple diffraction orders (grating lobes) are visible within the 5° x 9° angular beamwidth of the radiation pattern of the elementary grating coupler. (e) Detail of a thermally tunable beam coupler with a transparent monitor detector integrated at the output port. (f) Photograph of the photonic chip assembled on an PCB integrating the electronics for driving the thermal tuners and for the read-out of the on-chip detectors.

In this work, we show that a programmable mesh of MZIs can implement an adaptive multibeam receiver for a FSO communication link. The circuit can self-configure to simultaneously receive signals carried by spatially overlapped FSO beams sharing the same wavelength and polarization, provided that they are orthogonal in some way. Several examples are provided demonstrating that arbitrary pairs of orthogonal beams, sharing the same wavelength and polarization, can be separated with negligible mutual crosstalk; these examples include beams arriving from orthogonal directions (*direction-diversity*) as well as beams arriving from the same direction but shaped according to different orthogonal spatial modes (*mode-diversity*), and even when the two beams have undergone some mixing during propagation.

## II. MULTIBEAM FREE SPACE OPTICS RECEIVER

The programmable photonic circuit employed in this work consists of an integrated mesh of tunable beam splitters, which are realized by means of balanced Mach-Zehnder interferometers (MZIs). The topology of the circuit is shown in the schematic of Fig. 1(a) which includes $N = 2$ rows of cascaded MZIs [22]. On left side of the mesh, a 2D array of $M$ optical antennas ($M = 9$ in this particular device) is employed as an input/output interface between free-space optical beams and the guided modes of an array of single-mode optical waveguides. As discussed in Sec. III, in this work the optical antennas are implemented by using standard grating couplers typically used to couple the light with optical fibers; however, the presented results can be extended to integrated meshes terminated with arbitrary optical antennas, whose individual radiation diagram can be optimized for specific applications. On the right side of the mesh, two of the output waveguides, WG$n$ ($n$ = 1, 2), are used as output ports; the remaining 7 waveguide outputs are available for monitoring purposes.

An $M×N$ diagonal mesh can perform arbitrary linear transformations from an $M$-dimensional vector space to an $N$-

dimensional vector space [23] [24] . In this work this feature is exploited to implement a multibeam receiver for free-space optical beams. When an optical beam impinges from free space onto the photonic chip, the light is sampled by the 2D array of $M$ optical antennas and is coupled by the single-mode waveguides to the input of the programmable mesh. The light intensity in these $M$ waveguides can be coherently summed by configuring the first row of MZIs in such a way that the light in a first beam is entirely extracted out of WG1 with no residual power transmitted at the other output ports [22]. One can then simultaneously shine a second free-space optical beam, orthogonal to the first one, onto the 2D antenna array; this second beam can share the same wavelength and polarization as the first beam. As discussed in detail below, the orthogonality between the two beams can be the result of different arrival direction (see Sec. IV) or different spatial shape, typically referred to as mode orthogonality (see Sec. V). When the second beam is shone onto the photonic chip, its intensity is spatially sampled by the 2D antenna array and co-propagates with the first beam in the same $M$ single-mode waveguides, thus being apparently indistinguishable inside each waveguide. However, due to the original orthogonality of the two free-space beams, no portion of the second beam is transmitted to WG1 output port. The second row of MZIs stages can be used to coherently reconstruct the power of the second beam in the output port WG2. Generalizing the concept to a mesh with $M$ input waveguides and $N$ rows of MZIs, we can conclude that such a device can couple, reconstruct and separate $N$ orthogonal free-space beams, which are spatially sampled by $M$ optical antennas, and transmit them to $N$ single mode output waveguides, with arbitrary sorting order and no mutual optical crosstalk. This is the basic concept of the multibeam FSO receiver developed in this work.

If the propagation of the light is reversed, the mesh can operate as a multibeam transmitter [25], enabling us to map the light intensity carried by $N$ single mode waveguides into $N$ orthogonal free space beams. By tuning the integrated MZIs, both the amplitude and the phase of the light radiated by each element of the 2D optical antenna array can be controlled. In this way the shape and the direction of the far field beam can be modified. Note, incidentally, that these approaches change amplitudes by re-routing the light, not by absorbing or otherwise attenuating the beam, so there is no fundamental loss as relative amplitudes are adjusted. The number $M$ and the positions of the optical antennas set the effective spatial basis set [24] for the generation and collection of the free-space beams, and $M$ sets the number of degrees of freedom that we have for beam manipulation [26]. Even though in this work we restrict the analysis to the use of the programmable mesh as a multibeam receiver, backward propagation is exploited to understand better the behavior of the device under the conditions that optimize its performance.

## III. CHIP DESIGN AND FABRICATION

The programmable mesh of Fig. 1(a) was designed for operation in the 1550 nm wavelength range and was fabricated on a standard 220 nm SiP platform (AMF foundry). Fig 1(b) shows a top-view microscope picture of the entire chip, which has a footprint of 5.8 mm × 1.3 mm. All the waveguides of the circuit are single-mode channel waveguides with a width of 500 nm. As shown in the detail of Fig. 1(c), the 2D optical antenna array is made of $M = 9$ identical grating couplers, which are all aligned in the same direction and are arranged in a 3×3 square configuration. The grating couplers are designed to operate on transverse-electric (TE) polarized light. The emission angle with respect to the normal to the chip surface is 12°, while the radiation diagram has an angular beam width of 5° × 9° [26]. The center-to-center spacing between the grating couplers of the 2D optical antenna array is 49 μm (corresponding to about 32λ), this leading to the presence of several diffraction orders (grating lobes) in the far-field radiation pattern with minimum spacing of about 1.7°. For example, Fig. 1(d) shows the collimated far-field intensity profile (generated by a 50 mm Fourier transforming lens) and measured with a near-IR camera for a uniformly excited array (i.e., when all the elements radiate a light beam with same intensity and phase).

The nine waveguides connecting the grating couplers to the mesh share the same optical length in order to minimize the wavelength dependence of the multipath interferometer implemented by the mesh so the circuit can have the widest possible wavelength range of operation. All the 15 MZIs of the mesh (8 MZIs in the first row, 7 MZIs in the second row) are identical and are controlled by means of thermal tuners [see Fig. 1(e)]. Each MZI has two 3 dB directional couplers which are implemented by two 40 μm long waveguides spaced by 300 nm. Two thermal tuners made of TiN metal strips (2 μm x 80 μm) are integrated in each MZI stage (one in the lower input waveguide, one in the waveguide of the upper internal interferometer arm); these enable the control, respectively, of the relative phase shift between the optical fields at the output ports of the MZI and the amplitude split ratio of the MZI. Transparent photodetectors [27] are used to locally monitor the switching state of each MZI to implement automatic tuning and stabilization procedures.

The photonic chip was mounted on an electronic printed circuit board (PCB) [see Fig. 1(f)], housing also the drivers for the thermal tuners and the electronics (ASIC) for the read-out of the on-chip detectors [28]. The mesh self-configures and self-stabilizes by means of automated procedures, exploiting dithering signals for the thermal tuners [29] and thermal crosstalk mitigation strategies [30] [31]. Two parallel control systems were implemented for the control and calibration of each MZI row of the mesh independently. A turning mirror is positioned on top of the chip to steer the vertically emitted beam by the 2D array of grating couplers to the horizontal plane to facilitate the coupling with the free-space optical setup employed in the experiments presented in the following sections. Unless otherwise specified, all the experiments reported in the following sections are performed at a wavelength of 1550 nm.

## IV. DIRECTION-DIVERSITY RECEIVER

As a first example application, we show that the programmable mesh can operate as a *direction-diversity* receiver, that is, a multibeam receiver capable of individually detecting beams that simultaneously arrive from different directions. This concept is demonstrated here by considering the case of $N = 2$ beams, but this functionality can be generalized to $N$ beams arriving from $N$ directions utilizing a mesh with $N$ rows of MZIs. As shown in the scheme of Fig. 2(a), two free-space beams with identical gaussian shape, wavelength (1550 nm) and polarization status (TE polarization, to match the polarization sensitivity of the grating couplers) are shone from two different directions onto the 2D optical antenna array of the mesh. This implies that, in this case, orthogonality is given only by the direction of arrival of the beams, which is obtained by different position of the two transmitters, and we conveniently label the two beams as TX1 and TX2. The aim of this experiment is to demonstrate that the mesh can effectively separate the two beams TX1 and TX2 at the two output ports WG1 and WG2 (or vice versa) with negligible residual crosstalk.

Figure 2(b) shows a schematic of the experimental setup used to test the direction-diversity receiver. The free space optical beams are transmitted by two identical fiber coupled collimators (TX1 and TX2) that generate two gaussian beams with a waist of 1.12 mm. These two beams are imaged to the two ports of a 50/50 beam splitter (BS1) via a 4f system consisting of bi-convex lens with a focal length $f_3 = 250$ mm. Deliberately, the two beams are not overlapping in the beamsplitter plane (2.1 mm spacing), though they arrive overlapped at mesh inputs from different directions with a relative angle of 1.25°. The lens system between BS1 and the photonic chip is used to match the shape and size of the gaussian beams in center of BS1 to the collimated far field of the 2D optical antenna array. The system includes a pair of bi-convex lenses ($f_1 = 50$ mm) that are used in Fourier transforming condition to create the collimated far field of the 2D optical antenna array in the plane $P_1$, at the distance of 10 cm from the photonic chip, and the image plane of the 2D array at plane $P_2$ (20 cm away). Another bi-convex lens ($f_2 = 75$ mm) creates the required matching condition on BS1 at a distance of 350 mm from the chip. So, direction diversity implies that in the "far-field" planes ($P_1$ and $P_3$) the two gaussian beams are not spatially overlapped; however, in the 2D array plane on the chip surface (as well as in the image plane $P_1$) they are overlapped.

The beam impinging on the 2D optical antenna array from each source can be individually coupled to the desired output waveguide (WG1 or WG2) of the mesh through automatic tuning and stabilization algorithms applied to each row of the mesh [4], [29]. The mesh is self-configured by exploiting a local feedback control of each MZI, which is achieved by nulling the power of the beam at the monitor detectors in each row [see Fig. 1(a)] [26]. Although both beams co-propagate inside the waveguides of the mesh, they can be identified by the detectors provided that they are marked with suitable labels (pilot tones) superimposed as a shallow amplitude modulation at two different frequencies. In this experiment 500 Hz and 1kHz were chosen for this purpose. Upon tuning the rows of the mesh to extract each transmitted beam, identical optical powers from two collimators are measured at WG ports 1 and 2, which is reported as 0 dB normalized insertion loss in bar chart of Fig. 2(c). Notably, more than 25 dB optical crosstalk suppression, meaning TX1 in WG2 and TX2 in WG1, is measured and reported in the same bar chart. If we swap the mode sorting status, meaning coupling TX2 to WG1 and TX1 to WG2, the same level of optical isolation is observed.

To better understand the behavior of the mesh in the tuned state, we reversed the direction of the light propagation by injecting the light into WG1 and WG2 ports and measuring the far field pattern radiated by the 2D optical antenna array when the mesh is configured in the case of Fig. 2(c). The 2D far field profile shown in Fig. 2(d) is acquired using a near-infrared (NIR) camera focused on the plane $P_1$ after a 92/8 beam splitter (BS2). For clarity, we restrict the view to one period of the radiation pattern, which is zoomed in around the main lobe.

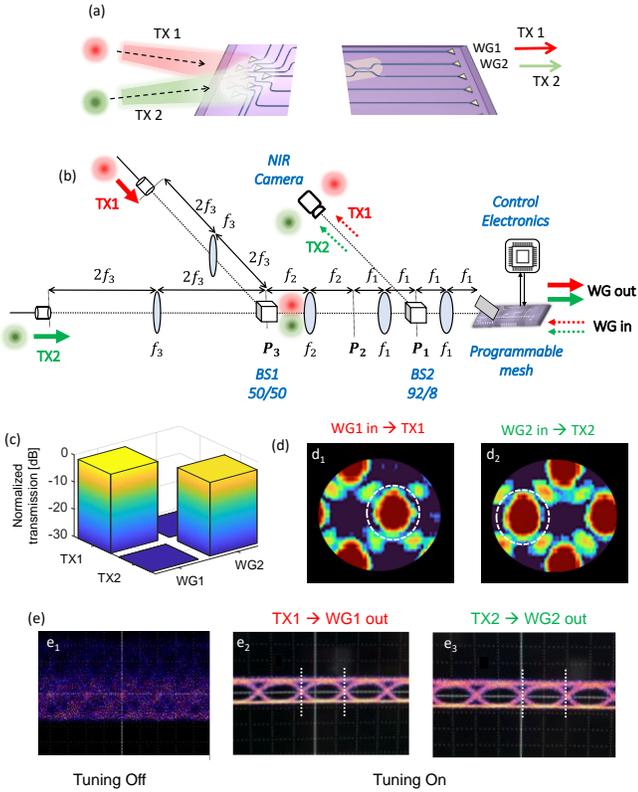

Figure 2: Direction-diversity receiver. (a) Schematic representation of two free space beams (TX1 and TX2), sharing the same wavelength and state of polarization, and arriving at the receiver from different directions. (b) Experimental setup employed for the demonstration of the direction-diversity receiver. (c) Bar chart showing the normalized insertion loss of the beams TX1 and TX2 at the output waveguides WG1 and WG2. (d) Backward far-field intensity pattern radiated by the 2D optical antenna array when the mesh is configured to couple beam TX1 to WG1 ($d_1$) and beam TX2 to WG2 ($d_2$); (e) measured eye diagrams of two intensity modulated 10 Gbit/s OOK signals transmitted by using the two beams TX1 and TX2: ($e_1$) when the mesh is not configured, the eye diagrams of TX1 and TX2 are severely overlapped; while after the mesh configuration both eyes, TX1 at port WG1 ($e_2$) and TX2 at port WG2 ($e_3$) are clearly open with neither evident distortion nor intersymbol interference.

Panels 2($d_1$) and ($d_2$) refer to the far-field radiated by the 2D optical antenna array when the light is injected from WG1 (transmission back to TX1) and from WG2 from (transmission back to TX2), respectively. It can be appreciated that in each case the position of the main lobe (highlighted by the dashed lines and indicating the direction of maximum radiation) coincides with the position of the null in the other case, which is consistent with the high rejection between the two beams shown in Fig. 2(c).

Such a low crosstalk level enables us to use the integrated mesh as a direction-diversity receiver in a FSO communication system, where the optical beams are employed to transmit two independent data channels. To this aim, the two beams TX1 and TX2 were used as carrier wavelengths for the transmission of independently modulated 10 Gbit/s on-off keying signals. The results of the transmission experiment are shown in Fig. 2(e). When the mesh is not set up in any particular way [panel ($e_1$)], that is when the thermal tuners of the MZIs are at arbitrary working points, the two data channels are randomly overlapped at output port WG1 and the measured eye diagram is completely closed. A similar result is observed at output port WG2 (not shown). In contrast, open eye diagrams are recorded when the mesh is tuned to extract the signal TX1 at output port WG1 ($e_2$) and the signal TX2 at output port WG2 ($e_3$); both eye diagrams then show no degradation with respect to the reference eye diagram of the individual channels.

## V. MODE-DIVERSITY RECEIVER

In principle, the coupling, separation and sorting of free-space beams can be operated by the programmable mesh on any set of orthogonal beams. As a second example, we consider two free-space beams, sharing the same wavelength (1550 nm) and state of polarization (TE), and coming from the same direction, yet being shaped according to different orthogonal spatial modes. In this case, the programable mesh receiving these two beams operates as a multibeam *mode-diversity* receiver. This situation is schematically shown in Fig. 3(a), where two free-space beams, labelled as Mode1 and Mode 2, simultaneously impinge on the 2D optical antenna array; the mesh is configured to separate them at the two output ports WG1 and WG2 (or *vice versa*). In the experiment described in the following, Mode 1 and Mode 2 refer to the fundamental Hermite-Gaussian mode HG00 and a higher order HG10-like mode, respectively.

In order to generate the two spatially overlapped orthogonal beams, the setup of Fig. 2(b) is slightly modified as shown in Fig. 3(b). In front of the fiber collimator TX2 a phase mask (Silica thin layer etched) is positioned that introduces a 0-π phase jump across the beam. As a result, the gaussian beam at the output of the fiber collimator (HG00) is transformed to a higher-order HG10-like mode. (It is not exactly a mode HG10 because the amplitude pattern does not have exactly that form; rather, a sudden π-phase shift is introduced between one side of the beam and the other on the vertical axis of the beam, giving a "two-bumped" beam with the same symmetry and phase behavior as an HG10 mode.) The second difference in the setup is that the positions of the two fiber collimators are optimized in such a way that the two modes are spatially overlapped in the far field planes ($P_1$ and $P_3$) and arrive on the 2D antenna array with the same direction.

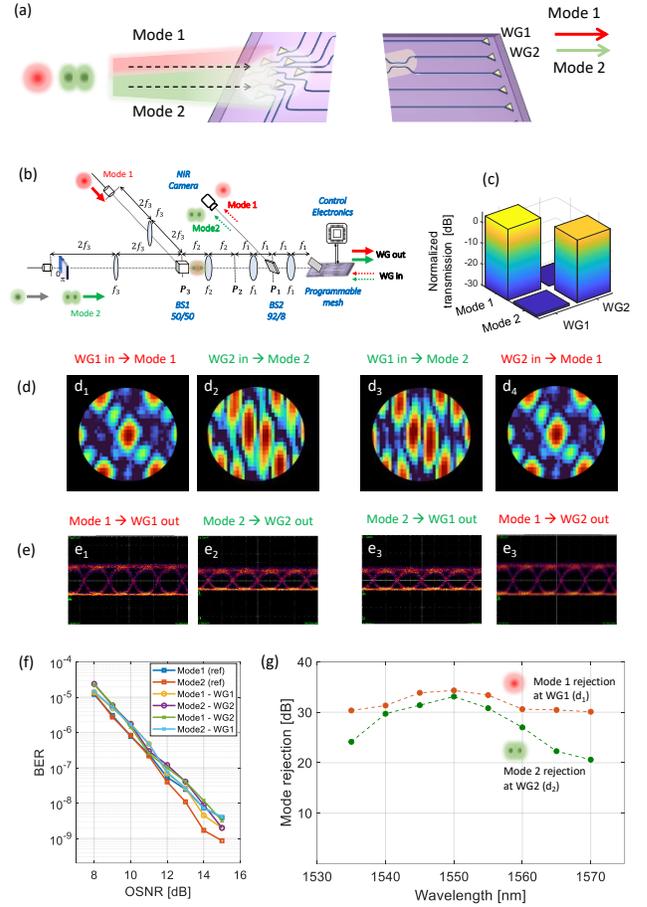

Figure 3: Mode-diversity receiver. (a) Schematic representation of two free-space modes (Mode 1 and Mode 2), sharing the same wavelength and state of polarization, and arriving at the receiver from the same direction. (b) Experimental setup employed for the demonstration of the mode-diversity receiver. In the reported experiment, Mode 1 and Mode 2 correspond to mode HG00 and mode HG10-like, respectively. (c) Bar chart showing the normalized insertion loss of the Mode 1 and Mode 2 at the output waveguides WG1 and WG2. (d) Backward far-field intensity pattern radiated by the 2D optical antenna array when the mesh is configured to couple Mode 2 to WG1 ($d_1$), Mode 1 to WG2 ($d_2$), Mode 1 to WG1 ($d_3$) and Mode 2 to WG2 ($d_2$); (e) measured eye diagrams of two intensity modulated 10 Gbit/s OOK signals transmitted by using Mode 1 and Mode 2 for the mesh configurations considered in (d). (f) BER measurements of 10 Gbit/s OOK channels simultaneously transmitted in free space on spatially overlapped modes (Mode 1 and Mode 2) and separated by the mesh. Blue and red squares indicate the reference BER measured when only one data channel is switched on. Negligible OSNR penalty is observed for any data channels sorted out by the mesh. (g) Wavelength dependence of the mutual mode rejection at the output ports of the mesh.

Adopting the same automated procedure described in Sec. IV for configuration of the circuit, the first MZI row of the mesh is lined up to maximize the coupling from one of the two modes (for instance Mode 1) to the output port WG1, which leads to nulling of the other mode (Mode 2) at this port. The bar chart of Fig. 3(c) shows the relative transmission of the two modes measured at both output ports after the configuration of the mesh. An intensity ratio of more than 30 dB between the extracted mode and the rejected mode is observed at both ports.

If mode sorting is swapped, that is if Mode 2 is coupled to WG1 and Mode 1 is coupled to WG2, the same level of isolation (more than 30 dB) is obtained.

By reversing the propagation direction of the light, the far-field profile of the beam generated by the 2D optical antenna array can be we observed for the various different configurations of the mesh. All the possible cases handled by the two-diagonal mesh are reported in Fig. 3(d) as measured by the NIR camera imaging the far-field plane $P_1$. For instance, let us consider the situation where the first MZI row of the mesh is configured in forward propagation to couple Mode 1 to output port WG1; in the reversed direction – that is, when WG1 is used as an input port – the far field radiated back by the 2D optical antenna array is well shaped as the fundamental HG00 mode [panel ($d_1$)]. In addition, if the second MZI row is configured to couple Mode 2 to WG2, the far field radiated back when WG2 is used as an input port is shaped like the HG10-like mode [panel ($d_2$)]. Panels $d_3$ and $d_4$ show the far-field pattern for the opposite coupling scenario. Notably, in all these cases the mesh automatically self-configures by simply minimizing the power of the relevant mode at each stage of the MZI rows, without any prior knowledge of the incoming beam shapes.

The performance of programmable mesh as a multibeam mode-diversity receiver was assessed by means of data channel transmission. Two intensity-modulated 10 Gbit/s OOK data streams were transmitted on the spatial and direction overlapped modes (HG00 and HG10-like) at the same carrier wavelength of 1550 nm and the same polarization state. The eye diagrams of the received signals, after the separation performed by the mesh, are shown in Fig. 3(e) for all the mesh configurations shown in Fig. 3(d). No degradation due to the residual mutual optical crosstalk from the interfering orthogonal mode can be observed. As a quantitative assessment of the effectiveness of the mode separation performed by the mesh, the bit error rate (BER) of the received channels was measured versus the optical signal to noise ratio (OSNR). The noise power in the OSNR is evaluated across a bandwidth equal to the bandwidth of the signal. Figure 3(f) shows the BER curves measured on the received data channels encoded onto the modes HG00 (Mode1) and HG10-like (Mode2). As reference curves, we measured the BER of the two modes (HG00 blue squares, HG10 red squares) when they are individually transmitted through the mesh to output port WG1 in the absence of the other mode. The other curves show the BER measured when both data channels are switched on and the modes are sorted out at the output ports WG1 and WG2 in all the four possible configurations. Thanks to the high optical crosstalk rejection between the separated modes, no significant OSNR penalty is observed in all the considered cases.

We also evaluated the wavelength range across which the mesh can guarantee a high isolation in the separation of the two modes. To this end, the carrier wavelength of the two modes was swept across a 35-nm-wide range from 1535 nm to 1570 nm. The width of this range is mainly limited by the wavelength-selective response of the grating couplers of the 2D optical antenna array. In the results reported in Fig. 3(f), for every wavelength considered, the mesh was configured to extract Mode 2 at output port WG1 and Mode 1 at output port WG2; these are the cases considered in the eye diagrams of panels ($d_3$) and ($d_3$) for the central wavelength of 1550 nm. The red curve shows that the intensity rejection of Mode 1 at port WG2 is higher than 30 dB across the entire wavelength range. Rejection is somewhat lower for Mode 2 at port WG1, yet is still higher than 20 dB across 35 nm.

## VI. MODE-MIXED RECEIVER

In this section we extend the concept of the mode-diversity receiver presented in Sec. V to other examples of spatially overlapped beam pairs that can be disentangled by the programmable mesh. In particular, we show that the mesh can still separate two orthogonal free-space beams even after they have propagated spatially overlapped through a mode mixing obstacle or a free-space path perturbation. Mathematically, this means that the linear transformation performed by the mode mixer, which maps the original orthogonal modes to another pair of orthogonal beams, can be inverted by the mesh.

Referring to the experimental setup of Fig. 3(b), in the far-field plane $P_1$ the fundamental mode HG00 (Mode 1) and the higher-order HG10-like mode (Mode 2) are spatially overlapped. If the phase mask in front of collimator TX2 is rotated -45 degrees, the axis of the Mode 2 in $P_1$ will be rotated accordingly, as shown in the schematic of Fig. 4(a). Now, we introduce in the plane $P_1$ (instead of BS2) a 0-$\pi$ phase mask which is rotated by 45 degrees with respect to the vertical axis. After passing through this mask, the fundamental mode HG00

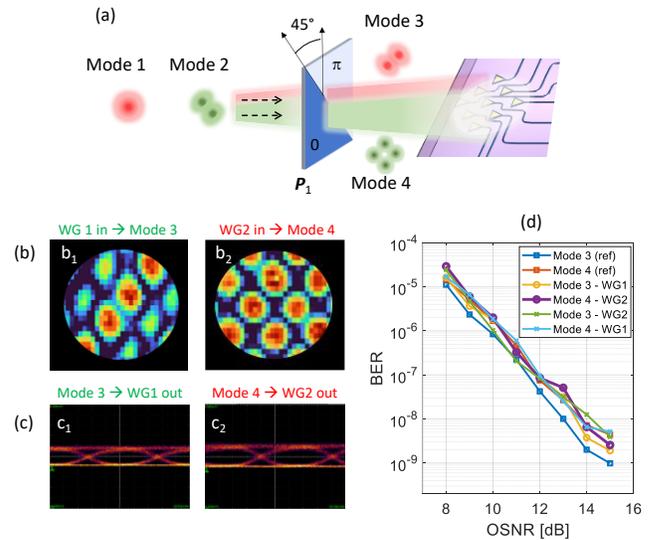

Figure 4: Mode separation after an arbitrary mode-mixing. (a) Schematic representation of two free space beams (Beam A and Beam B) that arrive on the receiver after an unknown linear transformation performed by a randomly oriented phase mask acting on two orthogonal modes (Mode 1 is HG00 and Mode 2 is 45-deg rotated HG10-like). (b) The shape of Beam A and Beam B can be identified by observing the backward far-field intensity pattern radiated by the 2D optical antenna array when the mesh is configured to couple Beam A to WG1 ($d_1$) and Beam B to WG2; (c) BER measurements of 10 Gbit/s OOK channels simultaneously transmitted on the overlapped Beams A and B and separated by the mesh. Blue and red squares indicate the reference BER of individually transmitted channels.

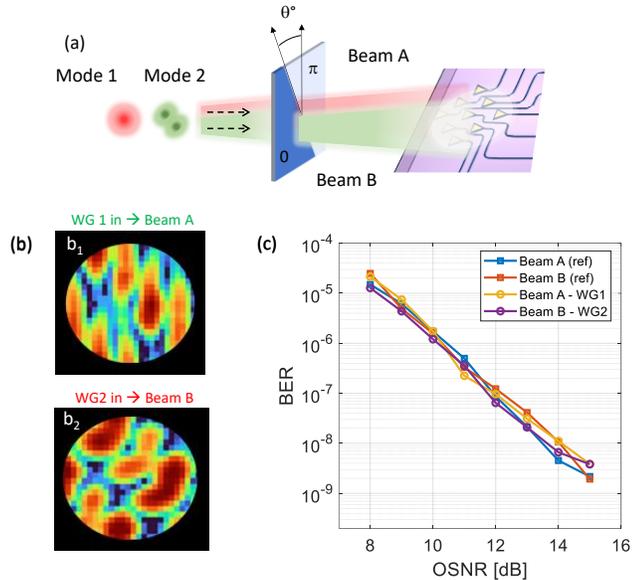

Figure 5: Mode separation after a controllable mode-mixing. (a) Schematic representation of two free-space modes (Mode 3 and Mode 4) that arrive on the receiver after a mode conversion performed by a 45-deg rotated phase mask. In the reported experiment, Mode 3 and Mode 4 correspond to 45-deg rotated HG10-like and HG11-like modes, respectively. (d) Backward far-field intensity pattern radiated by the 2D optical antenna array when the mesh is configured to couple Mode 3 to WG1 ($d_1$) and Mode 4 to WG2; (e) BER measurements of 10 Gbit/s OOK channels simultaneously transmitted in the free space on spatially overlapped modes (Mode 3 and Mode 4) and separated by the mesh. Blue and red squares indicate the reference BER of individually transmitted channels.

is converted to a 45-deg rotated HG10-like mode (Mode 3), while the higher-order mode HG10-like mode (-45-deg rotated) is transformed to a 45-deg rotated HG11-like (Mode 4). In other words, the phase masks perform a linear transformation between pairs of orthogonal modes, that is a mode conversion. From the point of view of the mesh, the operation needed to be performed to separate these new modes at the output ports WG1 and WG2 is conceptually identical to the one discussed in Sec. V. In fact, no a-priori information on the shape of the incoming beams is required and the self-configuring procedure to line up the MZI rows is exactly the same. Figure 4(b)-(d) show the results that are achieved when the mesh is tuned to extract Mode 3 at output port WG1 and Mode 4 at output port WG2. From the intensity ratio of the extracted modes at each out port, we observe more than 30 dB mutual isolation. Figures 4($b_1$) and ($b_2$) show the NIR camera acquisition of the far-field patterns that are radiated back by the mesh when the light is injected from port WG1 and WG2, respectively. A very good matching with the shape of the HG10-like mode and of the HG11-like mode (both rotated by 45 degree) is found. The good mode separation is confirmed when the mesh is used as a two-beam mode-diversity receiver in a data transmission link, where each mode carries a 10 Gbit/s intensity modulated OOK signal. Neither distortion nor inter-symbol interference effects are visible in the received eye diagrams of Fig. 4($c_1$)-($c_2$), which refer respectively to the data channel transmitted on Mode 3 extracted at output port WG1 and to the data channel transmitted on Mode 4 extracted at output port WG2. Figure

4(d) shows the BER curves versus OSNR measured on the data channels separated by the mode-diversity receiver. With respect to the reference curves, given by the BER curves of Mode 3 (blue squares) and Mode 4 (red squares) when they are individually transmitted in the absence of the other mode, no OSNR penalty is observed when both data channels are transmitted and they are sorted out by the mesh at the output ports WG1 and WG2. These results validate the effectiveness of the mesh to separate generic pairs of orthogonal modes emerging from a free-space mode converter.

As a final example, we show that the same functionality can be performed by the mesh even on orthogonal beams that do not belong to any mode family. To prove this concept, let us consider the case of Fig. 5(a) where the 0-$\pi$ phase mask positioned in plane $P_1$ of the experimental setup of Fig. 2(a) is rotated at an arbitrary angle. Presuming that the phase mask does not introduce any relevant loss, it produces an arbitrary mixing of the incoming modes, resulting in the generation of two beams, namely Beam A and beam B, that are still orthogonal, which no longer resemble any of the modes of the HG family (though they are still describable as some linear combination of HG-like modes). The shape of these beams is not known if the axis of the phase mask is unknown. Nonetheless, the configuration of the mesh to separate them can be operated as discussed in the previous examples without any a-priori information on the incoming beams. Figure 5(b) shows indeed that the mutual isolation between Beam A and Beam B is higher than 28 dB when they are separated at WG1 and WG2. If we want to know the shape of the two beams A and B, we can reverse the direction of propagation, injecting the light at ports WG1 and WG2 and looking at the far-field radiated by the 2D optical antenna array with the NIR camera. The field profiles of Beam A and Beam B are shown in Figs. 5($c_1$) and 5($c_2$) and, as expected, they exhibit an arbitrary shape that does not match any of the tabulated optical free space modes. Nonetheless, they are still orthogonal and they can be separated with extremely low mutual crosstalk, as confirmed by the BER measurements shown in Fig. 5(d).

Such modes are consistent with the general description of communications modes or mode-converter basis sets for arbitrary optical systems [24]; such modes need not correspond to any standard families, and need not be the same at transmitter and receiver, yet they still can have the key orthogonality properties of modes and independent communication channels.

### VII. CONCLUSION

In this work, we reported on an integrated adaptive multibeam receiver for FSO communications that is realized by means of a programmable silicon photonic mesh of thermally tunable MZIs. A 9x2 mesh ($M = 9$ input optical antennas, $N = 2$ output waveguides) was used to simultaneously establish two communication channels that are associated with spatially overlapped FSO beams sharing the same wavelength and polarization. The photonic circuit can be operated as either a direction-diversity receiver or a mode-diversity receiver, and can effectively separate pairs of orthogonal beams even after they have undergone mode mixing. In all the cases considered, the beams are separated and arbitrarily sorted at the two output ports with negligible mutual crosstalk (< -25 dB), allowing the use of the receiver in SDM systems, as demonstrated by data channel transmission at 10 Gbit/s. The optical bandwidth of the receiver, spanning across the extended telecom C-band (1530 nm – 1570 nm) allows its use in high data-rate systems as well as in wavelength-division multiplexing (WDM) – SDM communication links.

The automated control strategies used for the configuration of the MZI mesh, exploiting the implementation of local feedback loops in each MZI stage, enable scalability of the proposed architecture to circuits with a larger number of optical antennas, as well as to MZI meshes with a larger number of rows, which can handle more orthogonal beams. Moreover, the adaptive nature of the receiver would also allow the possibility of compensating for dynamic changes in the FSO link, caused by, for instance, moving obstacles or turbulence, so as to maintain the optimum communication link. Applications are also envisioned to more advanced FSO processing, including wave-front sensing, phase front mapping and reconstruction, beaming through scattering media and chip-to-chip FSO communications.

## VIII. Acknowledgements

This work was supported by the European Commission through the H2020 project SuperPixels (grant 829116). DABM acknowledges support from the Air Force Office of Scientific Research (AFOSR) under award number FA9550-17-1-0002. Part of this work was carried out at Polifab, the micro- and nanofabrication facility of Politecnico di Milano (https://www.polifab.polimi.it/). We thank Marco Sampietro for his support in the realization of the control electronics and Giorgia Benci for her contribution to the measurements reported in this work.